\documentclass[amsmath,amssymb,11pt,tightenlines,superscriptaddress,preprintnumbers, notitlepage,aps,prd,twocolumn,]{revtex4-1}

\pdfoutput=1 

\usepackage{mathtools}
\usepackage[utf8]{inputenc}
\usepackage[english]{babel}
\usepackage{amsmath}
\usepackage{graphicx}
\usepackage{epsfig}
\usepackage{slashed}
\usepackage{amssymb}
\usepackage{color}
\usepackage{tabulary}
\usepackage{multirow}
\usepackage[normalem]{ulem}
\definecolor{MyDarkBlue}{rgb}{0.1, 0.1, 0.8} 
\definecolor{SBlue}{rgb}{0.2, 0.4, 0.7} 
\definecolor{MyLightBlue}{rgb}{0.22,0.51,0.9}
\definecolor{MyGreen}{rgb}{0.0, 0.5, 0.0}
\definecolor{BrickRed}{rgb}{0.8, 0.25, 0.33}
\RequirePackage{hyperref}
\hypersetup{colorlinks, citecolor=SBlue,linkcolor=MyDarkBlue, urlcolor=BrickRed}
\makeatletter

\begin{document}
\preprint{FERMILAB-PUB-20-096-T}
\preprint{OSU-HEP-20-02}

\title{\Large Neutrino Non-Standard Interactions:\\  Complementarity Between LHC and Oscillation  Experiments}

\author{K.S. Babu}
\email[E-mail:]{babu@okstate.edu}
\affiliation{Department of Physics, Oklahoma State University, Stillwater, OK, 74078, USA}

\author{Dorival Gon\c{c}alves} 
\email[E-mail:]{dorival@okstate.edu}
\affiliation{Department of Physics, Oklahoma State University, Stillwater, OK, 74078, USA}

\author{Sudip Jana}
\email[E-mail:]{sudip.jana@mpi-hd.mpg.de}
\affiliation{Max-Planck-Institut f{\"u}r Kernphysik, Saupfercheckweg 1, 69117 Heidelberg, Germany}

\author{Pedro~A.~N.~Machado}   
\email[E-mail:]{pmachado@fnal.gov}
\affiliation{Fermi National Accelerator Laboratory, Batavia, IL, 60510, USA}

\begin{abstract}
We explore the complementarity between  LHC searches and neutrino experiments in probing neutrino non-standard interactions. 
Our study spans the theoretical frameworks of effective field theory, simplified model and an illustrative UV completion, highlighting the synergies and distinctive features in all cases. We show that besides constraining the allowed NSI parameter space, the LHC data can break important degeneracies present in oscillation experiments such as DUNE, while the latter play an important role in probing light and weakly coupled physics undetectable at the LHC.
\end{abstract}

\maketitle


\noindent{\bf I. Introduction --}
In spite of its vast success, the Standard Model (SM) does not shed any light on the origin of neutrino masses.  Overwhelming experimental evidence has shown that neutrinos of different flavors oscillate among one another, which cannot occur without small neutrino masses.  New physics is necessary for generating such masses, and thus the study of neutrinos offers a promising window for physics beyond the Standard Model.

The new physics associated with neutrino masses can lie in a vast range of energy scales: from  sub-GeV to the TeV region,  even reaching unification scales of order $10^{14}$~GeV.
This poses a phenomenological challenge that should be addressed with multiple experiments probing different energy scales, and in combination with suitable theoretical frameworks. 
For instance, neutrino Non-Standard Interactions (NSI) with matter can change the neutrino oscillation probabilities~\cite{Wolfenstein:1977ue}, where the momentum transfer is negligible, $q^2 \to 0$~(for a summary of current status of NSI see Ref.~\cite{Dev:2019anc}).
Therefore, the impact of new physics in oscillation experiments is well described by Effective Field Theory (EFT).
In contrast, at high energy colliders, the momentum transfer can be sizable, $q^2 \sim$~TeV$^2$, possibly leading to direct production of new states, where the  consistency of EFT may no longer be guaranteed~\cite{Davidson:2011kr,Berezhiani:2001rs,Friedland:2011za,Franzosi:2015wha,Choudhury:2018azm, Bischer:2019ttk}.
Here, a well suited framework should at least include the new degrees of freedom, as in simplified models, or  may display an even richer phenomenology, as in a UV complete scenario.

The purpose of this paper is to highlight the complementarity of neutrino experiments such as DUNE and collider searches in probing NSIs, across this multitude of frameworks: from the EFT to simplified models, and an illustrative UV completion. 
In Sec.~II we define EFT and simplified scenarios and use them to evaluate the LHC sensitivity to NSIs, in Sec.~III we discuss the complementarity between LHC 
and oscillation physics, in Sec.~IV we present an illustrative UV completion, and in Sec.~V we conclude.

\vspace{0.1cm}
\noindent{\bf II. From EFTs to Simplified Models --}
Neutrino oscillation probabilities are modified in a medium in the presence of NSI, which are generally parametrized in the EFT framework as:
\begin{equation}
\mathcal{L}_{\text{NSI}} = -2\sqrt{2}G_F \epsilon_{\alpha\beta}^{fY}\! \left(\bar{\nu}^\alpha\gamma_\mu \nu^\beta    \right) \left(\bar{f}\gamma^\mu P_Y f \right) \, ,
\label{eq:NSI}
\end{equation}
where $\epsilon_{\alpha\beta}^{fY}$  defines the strength of the $\nu-f$ interaction, $Y=L$ or $R$, $P_L$ $(P_R)$ is the left (right) 
chiral projector,  $\alpha,\beta\in \{e,\mu,\tau\}$, and $f = \{u, d, e\}$.

These new physics contributions can arise from higher dimensional operators that are invariant under the SM gauge symmetry. 
The dominant effects are expected to come from dimension six operators such as
\begin{equation}
\frac{1}{\Lambda^2} \left(\overline{L}_\alpha\gamma_\mu L_\beta    \right) \left(\bar{q}\gamma^\mu P_X q \right) \, ,
\label{eq:dimension-six}
\end{equation}
where $L$ is the lepton doublet and $X=L/R$. It follows that $\epsilon_{\alpha\beta} = - 1/(2 \sqrt{2}G_F \Lambda^2)$.  
One would expect prohibitively strong constrains on such operators from charged lepton flavor violation processes~\cite{Bergmann:1998ft}. However, flavor diagonal operators can avoid these constraints and  lead to observable NSI~\cite{Babu:2019mfe}.

Dimension 8 operators of the type \begin{equation}
    \frac{1}{\Lambda^4} \left(\overline{H L}_\alpha\gamma_\mu HL_\beta    \right) \left(\bar{q}\gamma^\mu P_X q \right)\,,
    \label{eq:dimension-eight}
\end{equation}
where $H$ is the Higgs doublet~\cite{Berezhiani:2001rs}, do not directly suffer from charged lepton flavor violation constraints, although there are other limits arising from non-unitarity of the PMNS matrix, oblique electroweak corrections, etc.~\cite{Gavela:2008ra}.  
If the mass of the mediator inducing Eq.~(\ref{eq:NSI}) is below the electroweak scale, charged lepton flavor violation constraints may even be absent~\cite{Babu:2017olk,Farzan:2015hkd}. 

While for oscillation experiments, we can safely take an agnostic approach to the origin the NSI operators and apply Eq.~\eqref{eq:NSI}, at the energy scales and couplings probed at the LHC, the validity of the EFT approach is no longer guaranteed. 
This discussion is similar to recent considerations about the interplay between dark matter (DM) searches at the LHC and low energy direct detection experiments~\cite{Fox:2011pm, Busoni:2013lha, Buchmueller:2013dya, Goncalves:2016iyg, Bell:2016ekl,Alanne:2017oqj}.  
Namely, while bounds from DM direct detection experiments on new physics can be interpreted in the EFT regime through operators like $\left(\bar\chi\gamma_\mu\chi\right) \left(\bar{q}\gamma^\mu q\right)/\Lambda^2$, the same does not hold true at the LHC, where the momentum transfers can go beyond the  validity of EFT.
Simplified models, in which the force mediator is dealt with explicitly, have been shown to be more appropriate for collider studies.
Adopting a  simplified model approach, we parametrize the NSI as
\begin{equation}
\mathcal{L}_{\text{NSI}}^{\text{Simp}} = \left(g_{\nu}^{\alpha\beta}\bar{\nu}_{\alpha}\gamma^\mu P_L\nu_{\beta}   + 
g_{qi}^Y  \bar{q}_i\gamma^\mu P_Y q_i  \right)Z'_\mu \,,
\label{eq:simp}
\end{equation}
where $Z'_\mu$ denotes the new force carrier with arbitrary mass $M_{Z'}$~\cite{Davidson:2011kr,Friedland:2011za}.

Manifestations of dark matter and non-standard neutrino interactions  at colliders would look quite similar, both involving large amount of missing transverse energy. A powerful probe of these interactions is the study of mono-jet signatures, in which  QCD initial state radiation leads to quark and gluon emission, $pp\rightarrow j+{E\!\!\!\!/}_{T} $ with $j=q,\bar{q},g$.
In this context, it has been shown that constraints from LHC 8 TeV data are more stringent for $M_{Z'}\gtrsim 100$~GeV
with the sensitivity reaching $\epsilon \lesssim 10^{-2}$~\cite{Friedland:2011za, Franzosi:2015wha}.

To estimate the current and high-luminosity LHC sensitivity to NSIs, we will recast the recent jets plus missing energy searches from ATLAS~\cite{Aaboud:2017phn} into the simplified model of Eq.~\eqref{eq:simp}. 
The limit $M_{Z'}\gg \sqrt{s}$ can be identified as the EFT in Eq.~\eqref{eq:NSI}. 
For our analysis, we  generate the signal sample ${pp\rightarrow \nu\bar{\nu}j}$ with {\sc MadGraph5aMC@NLO}~\cite{Alwall:2011uj,Alwall:2014hca}, simulating the hadronization and underlying event effects with {\sc Pythia8}~\cite{Sjostrand:2007gs}. Detector effects are simulated with the {\sc Delphes3} package~\cite{deFavereau:2013fsa}. 

Following the recent 13~TeV ATLAS mono-jet study~\cite{Aaboud:2017phn}, we define jets with the anti-$k_t$ jet algorithm and radius parameter $R=0.4$, $p_{Tj}>30$~GeV and $|\eta|<2.8$ via {\sc FastJet}~\cite{Cacciari:2011ma}. Events with identified electrons with $p_T>20$~GeV or muons $p_T>10$~GeV in the final state are vetoed. To suppress the $Z$+jets and $W$+jets backgrounds, the events are selected with $\slashed{E}_T>250$~GeV  recoiling against a leading jet with $p_{Tj1}>250$~GeV, $|\eta_{j1}|<2.4$, and azimuthal separation $\Delta\phi(j_1,\vec{p}_{T,miss})>0.4$. Events with more than four jets are vetoed.

Although flavor diagonal NSIs may interfere with the SM background $Z(\nu\bar{\nu})$+jets, we find it to be negligible in the region of interest for the LHC sensitivity.
Therefore, the diagonal and non-diagonal NSIs result in equivalent bounds at LHC.  
Note also that, in fixing the total width, the number of signal events is proportional to $\epsilon^2\equiv  (\sum_{\alpha,\beta}|\epsilon_{\alpha\beta}|^2)$ for both on-shell and off-shell production, and thus we use $\epsilon$ to quantify the LHC sensitivity to NSIs.
This eases the comparison to neutrino experiments without any further assumptions about the $g_\nu$ vs. $g_q$ ratio.

For concreteness, we will assume that neutrino NSIs arise in the simplified model~\eqref{eq:simp} as
\begin{equation}
\epsilon^{u}_{\alpha \beta} = \epsilon^{d}_{\alpha \beta} \equiv \epsilon_{\alpha \beta} = \frac{(g_\nu)_{\alpha \beta} g_{u,d}^V}{2 \sqrt{2} G_F M_{Z'}^2} \,,
\label{eq:epsilon}
\end{equation}
where $g_{u,d}^V = g_{u,d}^L + g_{u,d}^R$ is the quark vector current. We shall assume that $g_{qi}^A = 0$ and generation independent $Z'$ interactions with quarks $(g_{q}^V\coloneqq g_{qi}^V, \forall~i)$ for our numerical results, except in the context of an explicit model where these relations are not realized.

A major limitation of such searches is associated with the overwhelming SM backgrounds, ${pp\rightarrow Z(\nu\nu)j}$ and ${pp\rightarrow W(\ell\nu)j}$, which suffer from large theoretical uncertainties, including higher-order QCD and electroweak  corrections. 
Nevertheless, recently efforts have been made to improve the signal sensitivity in both experimental and theoretical fronts~\cite{Lindert:2017olm}. 
These include further exploration of background control regions and  state-of-the-art Monte Carlo simulations, resulting in  suppressed background uncertainties and augmented sensitivity to new physics. These advancements lead to suppressed systematic uncertainties that pave the way to stronger NSI constraints at the high-luminosity LHC.

\begin{figure}[t!]
    \centering
    \includegraphics[width=.5\textwidth]{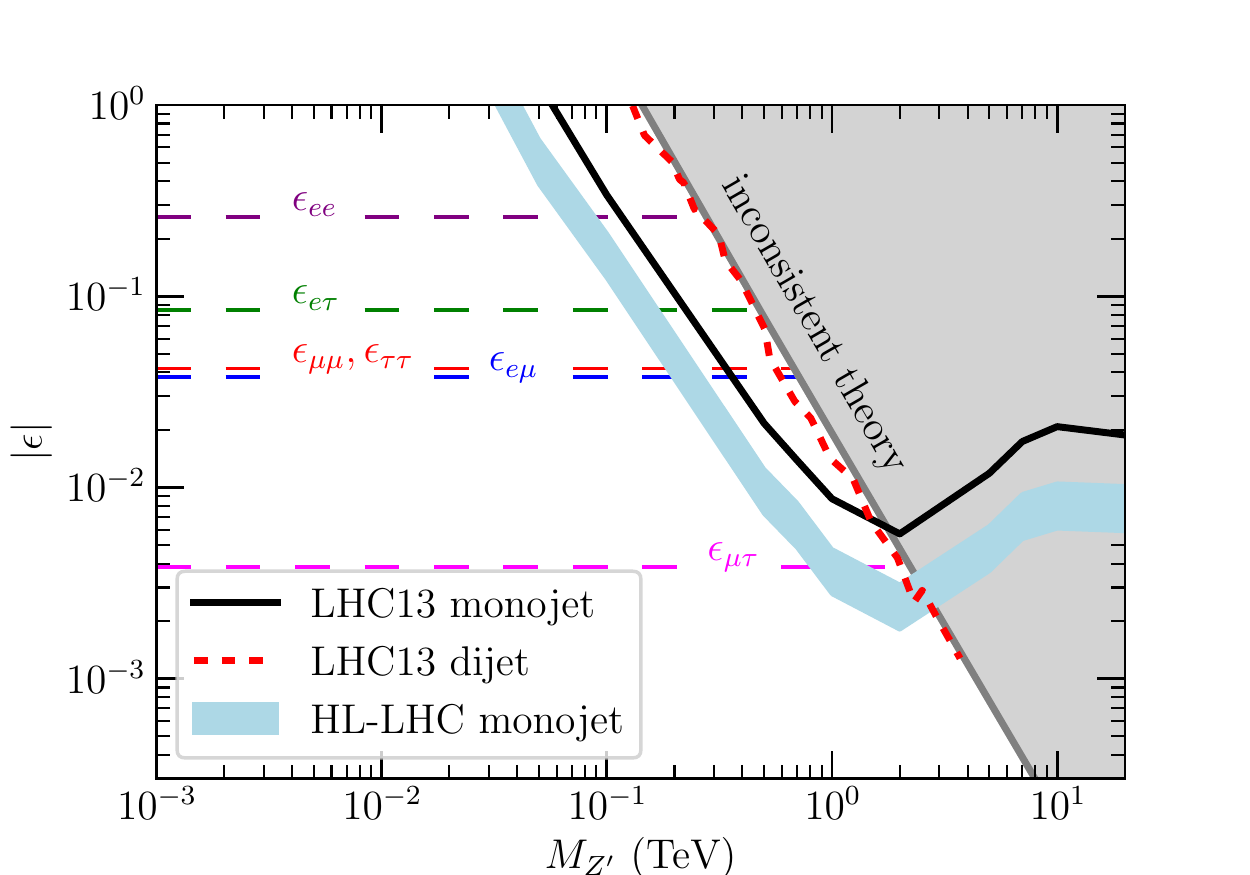}
    \caption{Constraints on  neutrino non-standard interactions from LHC data and neutrino experiments~\cite{Coloma:2019mbs} as a function of the mediator mass $M_{Z'}$, assuming $\epsilon_{\alpha\beta}\equiv\epsilon^u_{\alpha\beta}=\epsilon^d_{\alpha\beta}$.
    Note that LHC constraints are independent of neutrino flavor. We assumed $\Gamma_{Z'}/M_{Z'}=0.1$ here. 
    }
    \label{fig:nsi-constraint}
\end{figure}

In Fig.~\ref{fig:nsi-constraint}, we present the LHC sensitivity to NSIs at 95\% CL,  estimated for two LHC integrated luminosity scenarios:  36.1~fb$^{-1}$ and 3~ab$^{-1}$. 
For the lower luminosity case, ATLAS~\cite{Aaboud:2017phn} provides a limit on the signal events for ten signal regions. 
They differ by increasing $\slashed{E}_T$ thresholds. 
We define the NSI constraint  by the most sensitive region (black line). 
For the high-luminosity scenario, we obtain the backgrounds from ATLAS in the same ten $\slashed{E}_T$ thresholds. 
The  NSI sensitivity is obtained from the signal region that maximizes the significance
$\mathcal{S}=N_s/\sqrt{N_b+(\delta\sigma_b.N_b)^2}$,
where $N_s$ and  $N_b$ are the signal and background events, respectively, and $\delta\sigma_b$ is the background uncertainty. 
To evaluate the impact of systematic uncertainties at the high-luminosity LHC, the sensitivities are evaluated in two scenarios: $\delta\sigma_b=$1\% and 3\%.  
The future high-luminosity LHC sensitivity is shown as a blue region, in which the impact of such uncertainties is conveyed as a band. The resulting LHC constraint is maximal for mediators masses of order $M_{Z'}\sim 2$~TeV, reaching  $\epsilon\lesssim 2\times 10^{-3}$.
Flavor dependent bounds on NSI from neutrino experiments~\cite{Coloma:2019mbs} are shown as dashed lines and will be discussed in the following section. 

In addition to the jets plus missing energy searches, there are other relevant LHC constraints to this simplified model arising from di-jet resonance searches. 
We use a combination of these limits from ATLAS and CMS~\cite{Sirunyan:2018wcm,Sirunyan:2018xlo,CMS-PAS-EXO-19-012,Khachatryan:2016ecr,Sirunyan:2018pas,CMS-PAS-EXO-18-012,Sirunyan:2019sgo,Sirunyan:2018ryr}.
To set the most conservative bound from these data sets, we assume that the coupling of the resonance to neutrinos, $g_\nu$, saturates the chosen width $\Gamma_{Z'}/m_{Z'}=0.1$.
For this choice, the value of $g_\nu$ turns out to be always below 3.
These searches are complementary to mono-jets displaying significant sensitivities at large resonance  masses, $m_{Z'}\gtrsim 2$~TeV, see Fig.~\ref{fig:nsi-constraint}.

We can identify the EFT regime for the LHC when the mass of the mediator is much above the scale of the process involved. This can clearly be seen in Fig.~\ref{fig:nsi-constraint}, as the bound on $\epsilon$ does not change for mediator masses above $\sim 5$~TeV.
In addition, a robust argument can be made to estimate the validity of this EFT at the LHC.  
For any fixed ratio $\Gamma_{Z'}/M_{Z'}$, we can write the following inequality
\begin{equation}
|\epsilon| \leq \frac{ \sqrt{3} \pi}{\sqrt{N} G_F M_{Z'}^2} \frac{\Gamma_{Z'}}{M_{Z'}}~.
\label{theory}
\end{equation}
This constraint originates from the fact that the total width of the $Z'$ should be larger than the partial widths to $q_i\bar{q}_i$ and $\nu\bar{\nu}$: $\Gamma_{Z'} \geq M_{Z'}/(24 \pi)(g_\nu^2 + 3N \{(g^V_u)^2  + (g^V_d)^2\})$, where we assumed decay to a single neutrino flavor. 
$N$ here is the number of quark flavors below the threshold for $Z' \rightarrow \overline{f} f$ decay, with each flavor multiplied by a respective phase space factor.  
The expression for the partial decay width of $Z' \rightarrow \overline{f}f$ is given in the Supplemental Material, from which we see that the phase space factor for quark flavor $f$ is $P_f=\left(1-\frac{2m_f^2}{M_{Z'}^2}\right) \sqrt{1- \frac{4 m_f^2}{M_{Z'}^2}}$ (with axial vector coupling set to zero).  
Considering non-zero axial couplings would make the constraint more stringent. Thus, above $t\overline{t}$ threshold, $N = 5 + P_t$.  
Assuming $g^V_u = g^V_d$ leads to the ``inconsistent region'' (gray shaded) in Fig.~\ref{fig:nsi-constraint}.
Considering narrower $Z'$ makes the constraint stronger, while broader $Z'$ implies non-perturbativity ($\Gamma_{Z'}$ greater than roughly half $M_{Z'}$).
Therefore, traditional EFT analyses at the LHC using four-fermion operators like Eq.~\eqref{eq:NSI} will typically not be valid, at least having simple/minimal UV completions in mind.

\vspace{0.1cm}
\noindent{\bf III. Complementarity between LHC and neutrino experiments --} Differently from the LHC, the effects of NSIs in neutrino oscillations strongly depend on the flavor structure of the NSI and the oscillation channel being studied.

In Fig.~\ref{fig:nsi-constraint}, we show the limits on $|\epsilon_{\alpha\beta}|$ for each flavor combination, derived from the global fit~\cite{Coloma:2019mbs} using neutrino oscillation and COHERENT data (see also Ref.~\cite{Dutta:2020che}), where all other NSI parameters were marginalized over.
We have combined the limits on $\epsilon^u_{\alpha\beta}$ and $\epsilon^d_{\alpha\beta}$ at 95\% CL  in \cite{Coloma:2019mbs} in quadrature, $\sigma_{\alpha\beta}=[(\sigma_{\alpha\beta}^u)^{-2}+(\sigma_{\alpha\beta}^d)^{-2}]^{-1/2}$.
These constraints on NSI parameters span two orders of magnitude, showing the strong dependency on the flavor structure of the NSI.

The flavor dependence of NSIs on neutrino oscillations goes beyond different sensitivities.
The effects of different NSIs and/or variations of the standard oscillation parameters can, in some cases, compensate each other and lead to well known \textit{degeneracies}.  
Disentangling those is a difficult task at neutrino facilities.
In contrast, the mono-jet  signal at the LHC, $pp\rightarrow \bar{\nu}_\alpha\nu_{\beta}j$, does not distinguish between different choices of $(\alpha,\beta)$; {\it i.e.}, they all lead to the same observables. 
Besides constraining the currently allowed NSI parameter space, this feature can be further exploited to break  relevant degeneracies. 

\begin{figure}[t!]
    \centering
    \includegraphics[width=.5\textwidth]{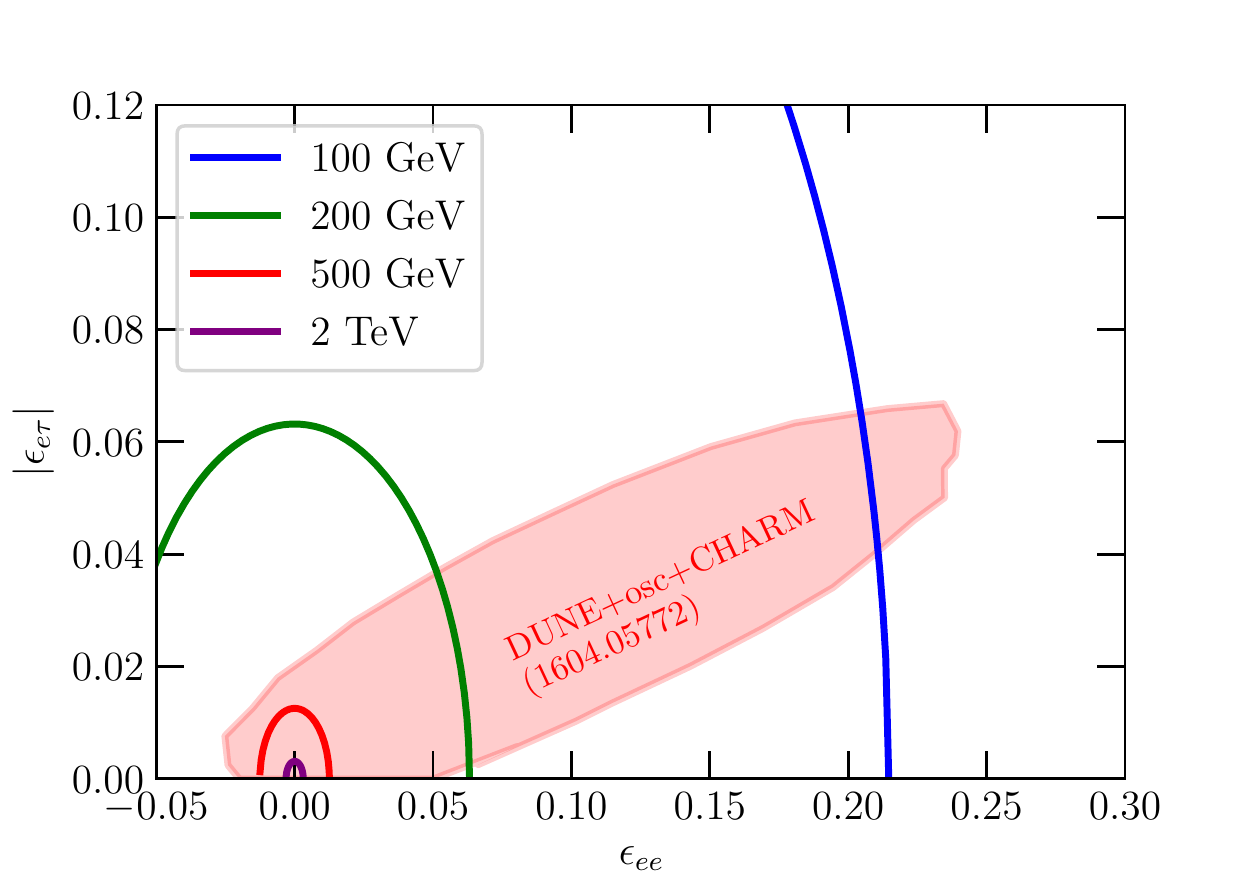}
    \caption{Constraints on  neutrino non-standard  interactions from LHC data, for different  mediator masses as labeled, with $\Gamma_{Z'}/M_{Z'}=0.1$, assuming $\epsilon_{\alpha\beta}\equiv\epsilon^u_{\alpha\beta}=\epsilon^d_{\alpha\beta}$.  A fit to simulated data for DUNE from Ref~.~\cite{Coloma:2016gei} is also included (pink shaded region). 
    }
    \label{fig:nsi-degeneracy1}
\end{figure}

\begin{figure}[t!]
    \centering
    \includegraphics[width=0.5\textwidth]{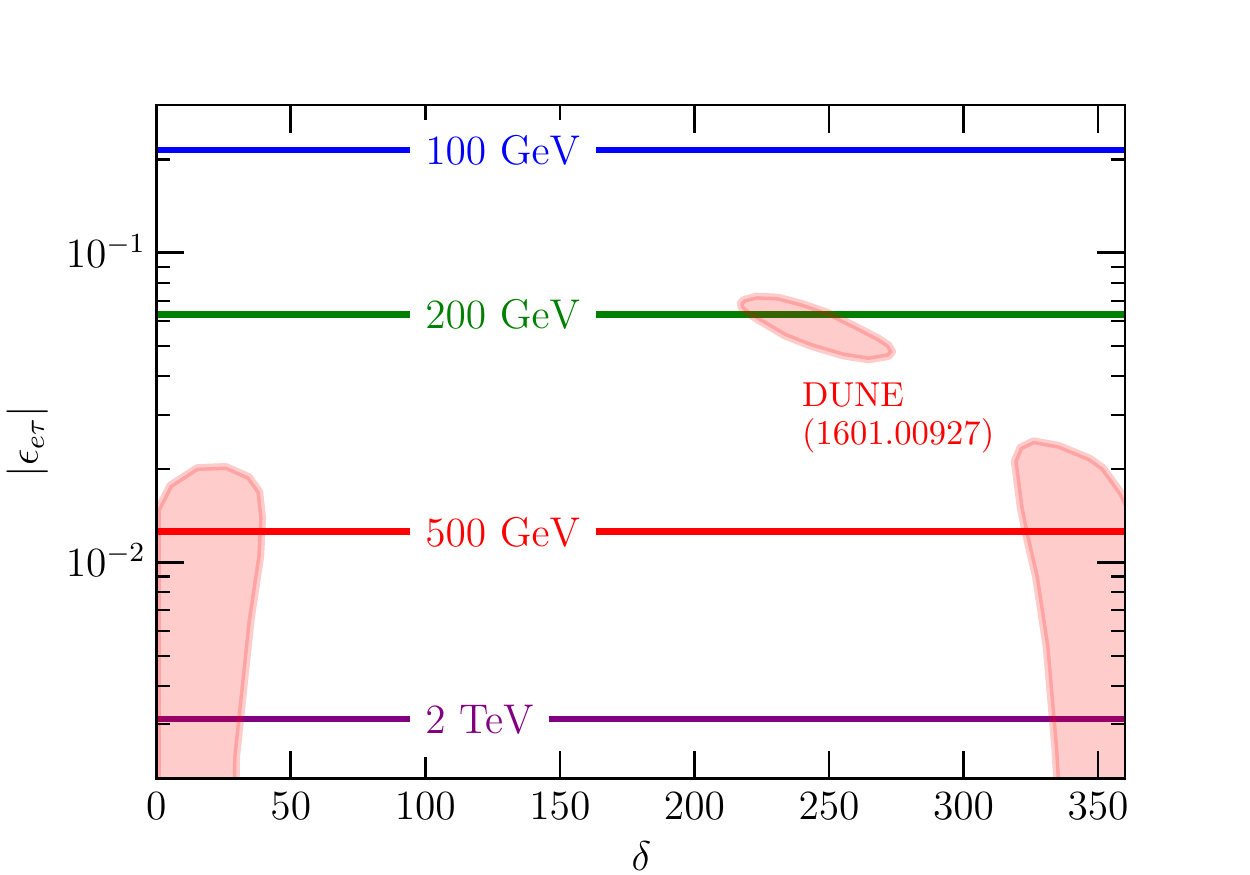} 
    \caption{An example of degeneracy in the $(\epsilon_{e\tau},\delta)$ plane, where $\delta$ is the standard $CP$ violating phase and $\epsilon_{e\tau}\equiv\epsilon^u_{e\tau}=\epsilon^d_{e\tau}$, from 
    future DUNE data 
    taken from Ref.~\cite{Liao:2016hsa} (pink shaded regions are allowed, assuming $\epsilon_{e\tau}=\delta=0$ as  null hypothesis). We overlay the LHC sensitivity to this NSI parameter for several mediator masses, as labeled, assuming  $\Gamma_{Z'}/M_{Z'}=0.1$.
    \label{fig:nsi-degeneracy2}}
\end{figure}

To make this point  manifest, we present two examples of degeneracy breaking in the following.
In Fig.~\ref{fig:nsi-degeneracy1}, we show the 95\% CL bounds on $|\epsilon_{ee}|$ vs. $|\epsilon_{e\tau}|$ from Ref.~\cite{Coloma:2016gei}, obtained by combining current oscillation and scattering data with future DUNE sensitivity. 
In Fig.~\ref{fig:nsi-degeneracy2}, we show the 95\% CL bounds on $\epsilon_{e\tau}$ vs. the usual $CP$ violating phase $\delta$ for 
the future DUNE experiment from Ref.~\cite{Liao:2016hsa}.
Here, $\nu_\mu\to\nu_e$ and $\bar\nu_\mu\to\bar\nu_e$ oscillation channels, driven by the smallest mixing parameter, $\sin^2 2\theta_{13}$, are crucial to constrain these parameters. 
These oscillations are significantly  affected by matter effects, and channels like $\nu_\mu\to\nu_\tau$, $\nu_e\to\nu_\mu$ are much harder to study due to experimental limitations.
The allowed regions from neutrino experiments are shown in pink shaded, while the LHC bound depends on the mediator mass and is depicted as colored lines, as indicated in the figures. 
It should be noted that, even with the future DUNE experiment, several degeneracies will remain unsolved by oscillation measurements, but could in principle be unravelled by LHC data.

Therefore, several important complementary aspects between LHC and neutrino experiments can already be identified.
The LHC sensitivity displays a strong dependence on the mediator mass, but it is free of parameter degeneracies.
Neutrino oscillation measurements, on the other hand, exhibit the opposite behavior: significant degeneracies and no mediator mass dependence.

On top of that, there is  another complementary aspect that cannot be seen from the figures presented so far.
The matter potential induced when neutrinos travel through a medium is not affected by a diagonal, universal contribution (as this just induces an overall phase shift on the neutrino state). 
On the other hand, LHC data is sensitive to each and all NSI parameters independently.
Note also that neutrino oscillations are not sensitive to axial interactions, while LHC data is sensitive to both vector and axial new physics contributions.
All these features show the synergies between oscillation measurements and collider data on probing new physics in the neutrino sector.

\begin{figure}[t!]
    \centering
    \includegraphics[width=.5\textwidth]{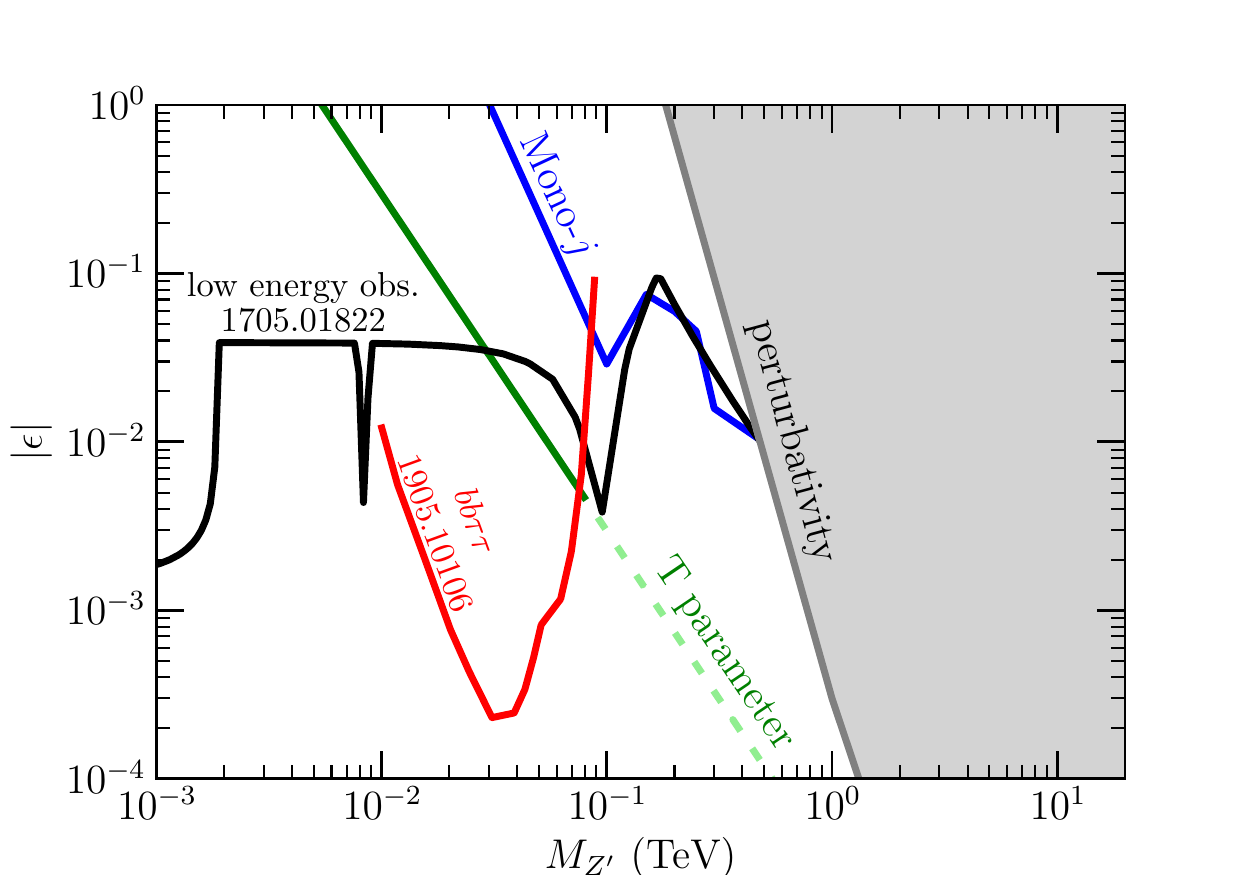}
    \caption{Sensitivities to a specific UV complete model realization of non-standard interactions where the $B-L$ number of the third family is gauged~\cite{Babu:2017olk}.
    This includes current low energy observables (black), electroweak $T$ parameter (green solid and dotted, see text), future high-luminosity LHC searches for $bb\tau\tau$ final states (red, adapted from Ref.~\cite{Elahi:2019drj}), and  LHC mono-jet searches (blue).
    \label{fig:UV-completion}}
\end{figure}

\vspace{0.1cm}
\noindent {\bf IV. Towards a UV complete scenario --} 
Any UV complete model of neutrino NSI is expected to provide a more extensive phenomenology, especially since neutrinos are in the same $SU(2)_L$ doublet as charged leptons. 
This further enhances the synergies between LHC and oscillation experiment, as we shall demonstrate in an illustrative UV completion. 
We show in Fig.~\ref{fig:UV-completion} the constraints on the UV complete model of Ref.~\cite{Babu:2017olk} (see Ref.~\cite{Han:2019zkz} for other constructions with leptonic signals).
In this UV completion the $B-L$ number is  gauged, but only for the third family. 
This leads to a new gauge boson that couples more strongly with the third family fermions and results in nonzero  $\epsilon_{\tau\tau}$. 
In Ref.~\cite{Babu:2017olk} the entire model was specified, and an extensive list of constraints were derived from low energy observables such as neutrino oscillation, $D-\bar D$ mixing, $\Upsilon$ and $B$ decays, atomic parity violation, and others. In the Supplemental Material, we provide a  description of the model, a few relevant equations, and updated constraints which are valid even for $M_{Z'}>M_Z$.

Assuming $\tan\beta=0.5$ (see Ref.~\cite{Babu:2017olk}), we compare these constraints (black line) with a dedicated $bb\tau\tau$ search~\cite{Elahi:2019drj} (red line), electroweak precision $T$ parameter~\cite{Tanabashi:2018oca} (green solid and dotted line) and the high-luminosity LHC jets plus missing energy constraint of Sec.~II (blue line).
The latter displays a distinctive profile due to enhanced  light flavors initial state contributions for $M_{Z'}\sim M_Z$ and  $b$-initiated at  $M_{Z'}\gg M_Z$.
If $M_Z<M_{Z'}$, the contribution to the $T$ parameter can in principle be compensated by those stemming from the scalar sector, making this constraint model dependent (indicated by the dotted green line). 
Finally, we also show the region in which the theory becomes non-perturbative, corresponding to the new gauge coupling being larger than 2.

Low energy constraints, dedicated LHC searches, and missing energy signatures provide strong constraints for different masses of the mediator.
For masses below about 10~GeV, low energy observables tend to dominate. 
In the  intermediate regime $10-100$~GeV, dedicated searches for visible signatures at the LHC become more relevant.
Finally, from $0.1-1$~TeV LHC mono-jet searches, low energy observables and electroweak precision observables (up to the $T$ parameter model dependence) play the leading role.
This makes manifest the complementarities among collider data, oscillation measurements, and other low energy observables.

\noindent{\bf V. Conclusion --}
We have explored the complementarity  between neutrino experiments and LHC searches in probing neutrino non-standard interactions. 
Our analysis covers the full span of theory frameworks: effective field theories, simplified models, and an illustrative UV completion. 
We have shown that the present and high-luminosity LHC sensitivities to NSIs display relevant synergies to oscillation results. Namely, $i)$~the breakdown of \emph{degeneracies} among NSI and oscillation parameters, and $ii)$ sensitivity to new phenomenological signatures at the LHC.
As a by-product of our analysis, we have shown that the use of EFT at the LHC in estimating sensitivity to NSIs is not generally theoretically consistent.

\vspace{-0.19in}
\section*{Supplemental Material}
\vspace{-0.13in}
\subsection{Decay width for $Z'$}
\vspace{-0.13in}
If the $Z'$ couplings to fermions  is
\begin{equation}
L_{Z'\overline{f}f} = g_L \overline{f}_L \gamma^\mu f_L Z'_\mu + g_R \overline{f}_R \gamma^\mu f_R Z'_\mu~,
\end{equation}
then the partial decay width of $Z'$ into fermion pairs is given by
\begin{eqnarray}
    \Gamma(Z' \rightarrow \overline{f}f) &=& \frac{N_c M_{Z'}}{24 \pi} \left[ (g_L^2+g_R^2)\left(1-\frac{m_f^2}{M_{Z'}^2}\right) \right. \nonumber \\  &+& \left. 6 g_L g_R\frac{m_f^2}{M_{Z'}^2}\right]\sqrt{1- \frac{4 m_f^2}{M_{Z'}^2}}, \label{width}
\end{eqnarray}
where $N_c$ stands for color degrees of freedom.  For our case, with purely vector couplings to quarks, we have in Eq. (\ref{width}) $g_L^q = g_R^q = g^q_V$, and $g_L^\nu = g_\nu$ for neutrino along with $g_R^\nu = 0$.
\vspace{-0.13in}
\subsection{UV complete model: expressions and corrected constraints}
\vspace{-0.13in}
The UV complete model from Ref.~\cite{Babu:2017olk}, used in Section~IV to exemplify the complementarity between LHC data and other experiments, can be summarized as follows.
The $B-L$ number of third family fermions is promoted to a $U(1)_{B-L}^{(3)}$ gauge symmetry.
To generate CKM mixing and to avoid a Goldstone boson, two doublets charged under the new $U(1)_{B-L}^{(3)}$ have to be added to the particle spectrum: a singlet and a doublet of $SU(2)_L$.
The ratio between vacuum expectation values of the new doublet to the Higgs is defined as $\tan\beta\equiv v_2/v_1$.

\begin{figure*}[ht!]
    \centering
    \includegraphics[width=0.45\textwidth]{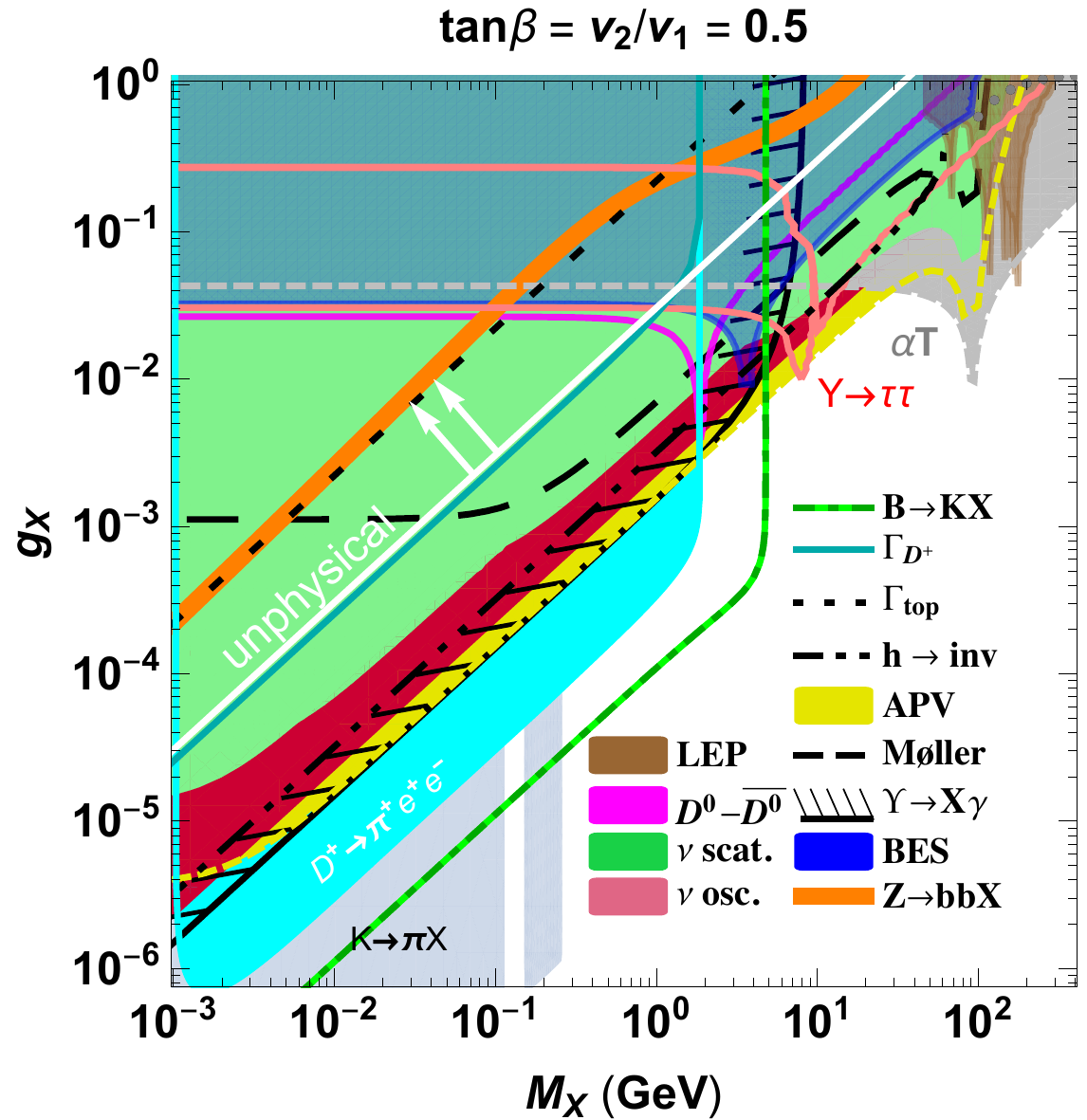} \includegraphics[width=0.45\textwidth]{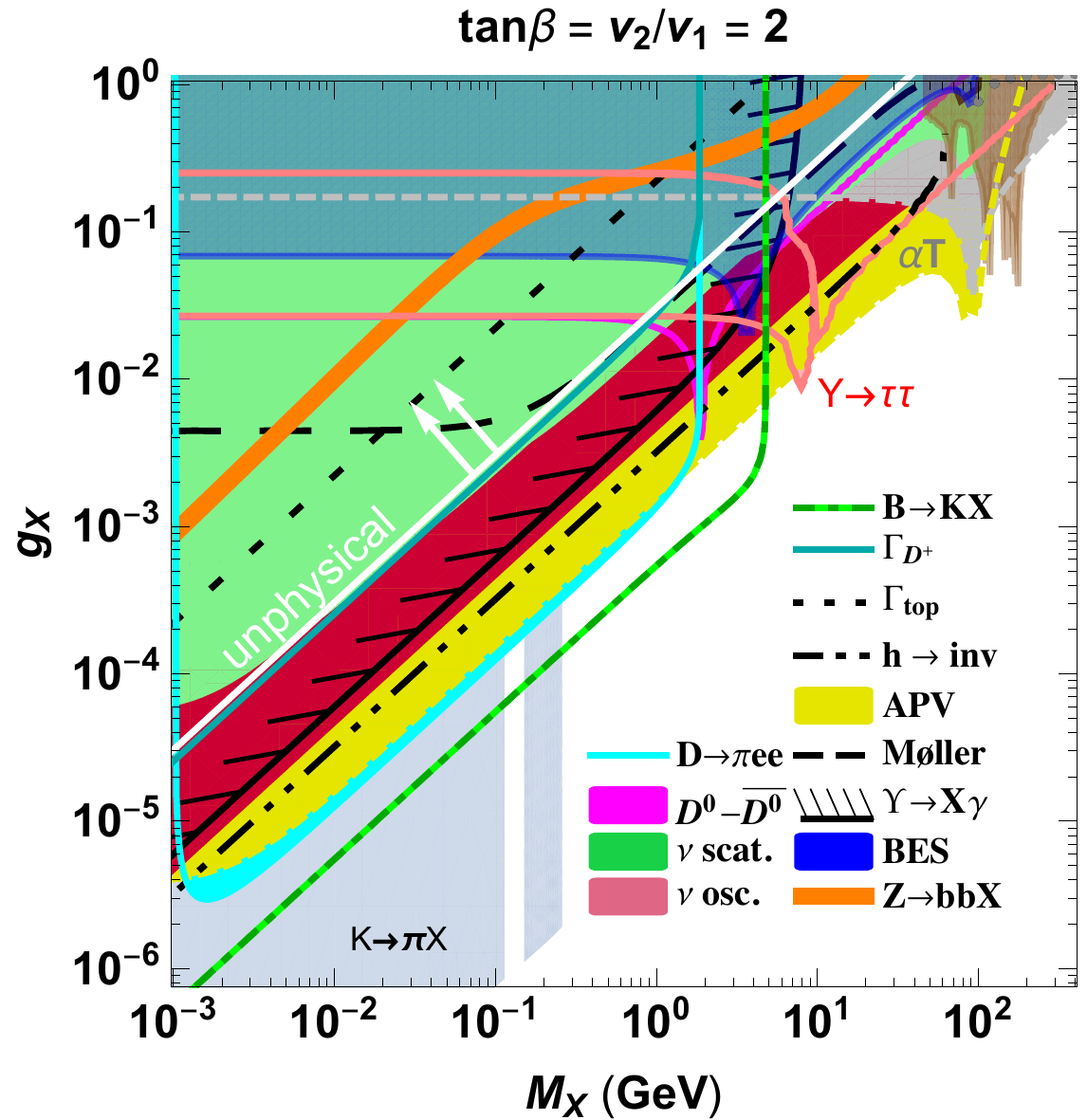} 
    \includegraphics[width=0.45\textwidth]{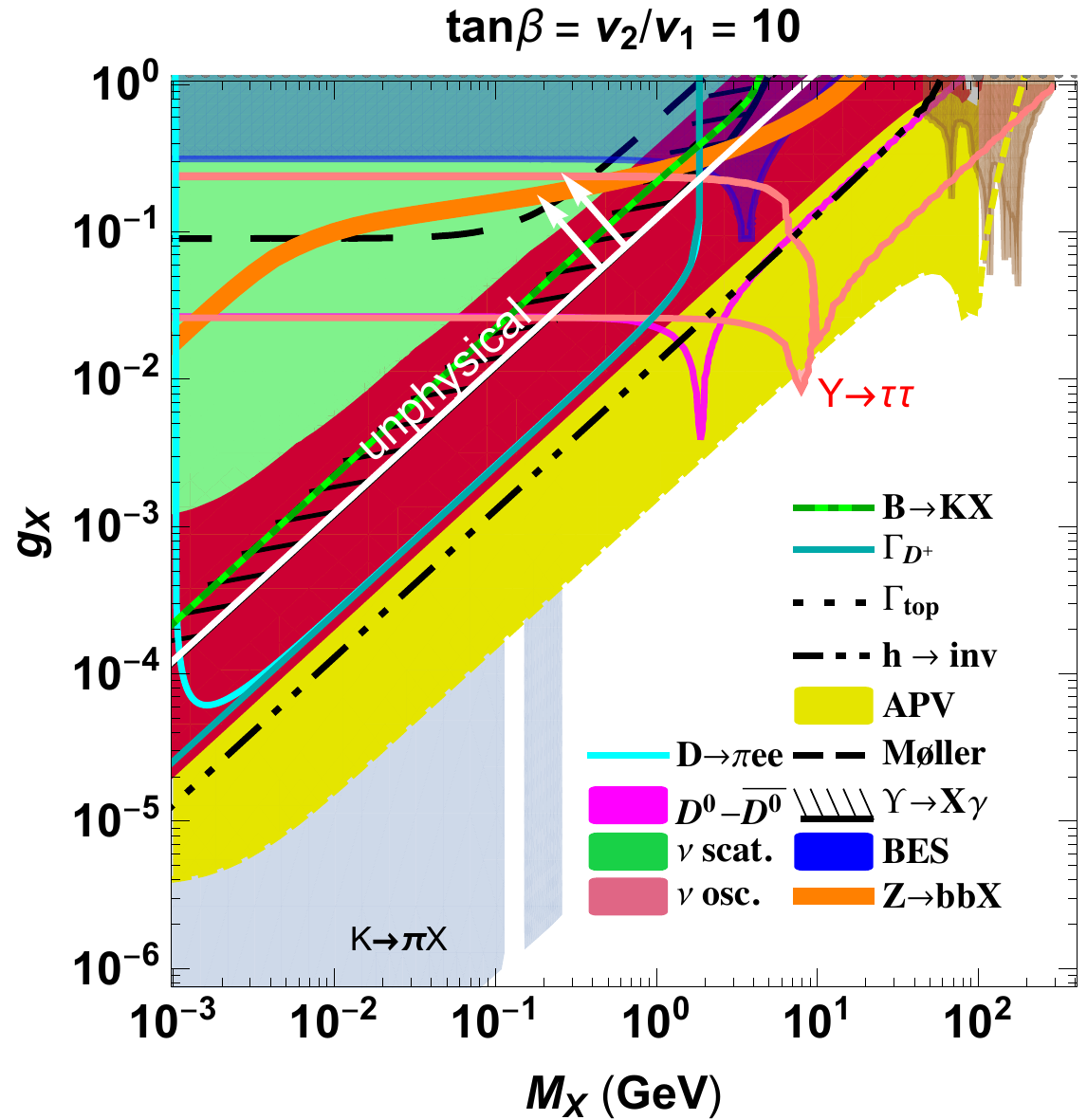} 
    \includegraphics[width=0.45\textwidth]{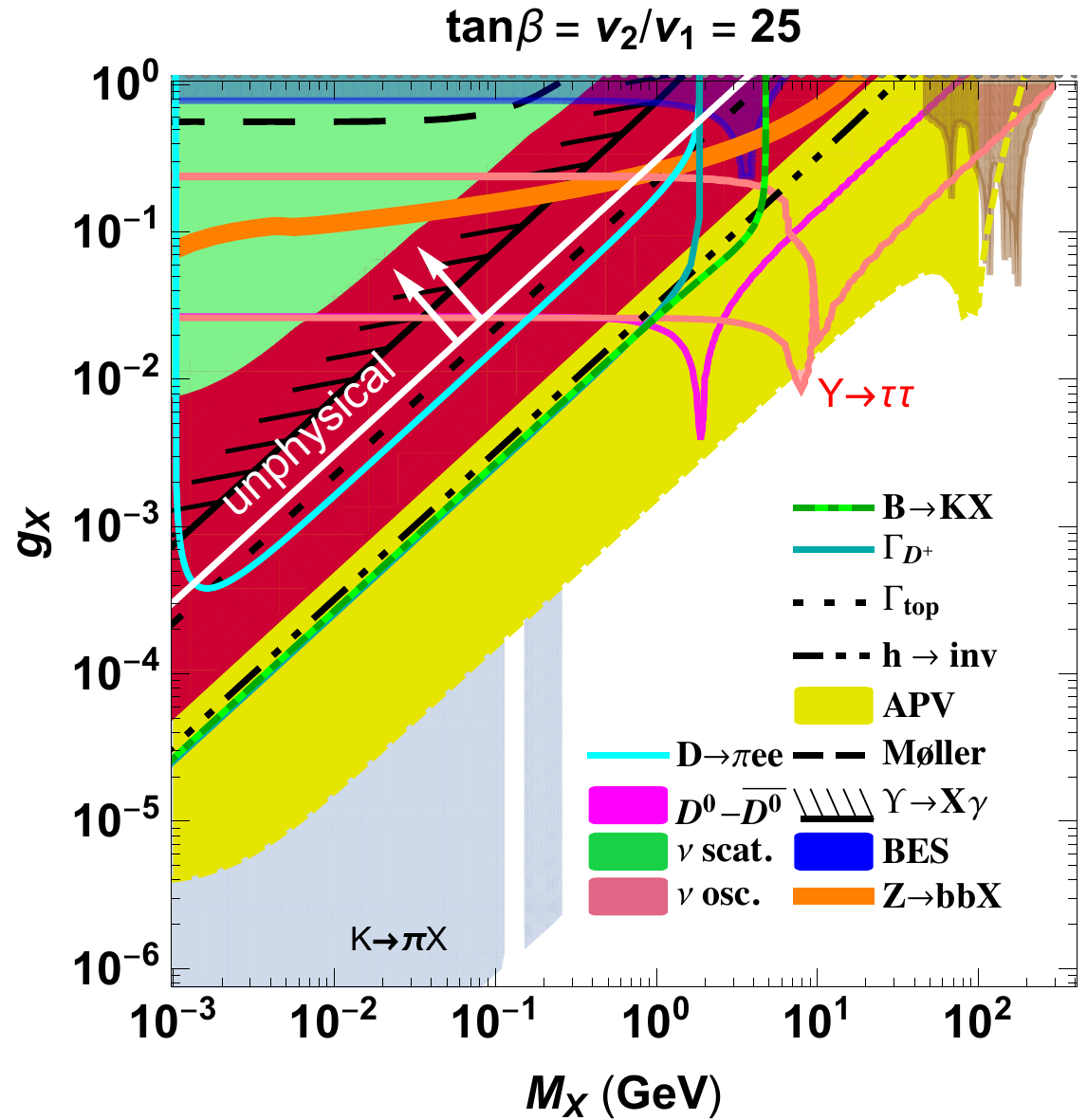} 
    \caption{Corrected constraints from Ref.~\cite{Babu:2017olk} in the plane $M_{X}$ versus $g_{X}$ (gauge coupling) for various values of $\tan\beta$, as indicated. The constraints come from the following experiments or observations: $B\to K+$invisible (dark/light green); decay width of $D^+$ meson (dark cyan) and top quark (dotted line); Higgs invisible decays (dot-dashed line); atomic parity violation (APV, yellow); M\"oller scattering (dashed line); $\Upsilon\to\gamma+$invisible (hatched region); BES (blue region); $Z\to bb+$invisible (orange thick line); neutrino oscillations (red region) and scattering (green region); $D^+\to\pi^+e^+e^-$ (cyan);  $D-\bar{D}$ oscillation (magenta); electroweak $T$ parameter (gray region); LEP data (brown region); $K^+\to\pi^++$invisible (light blue region); and $\Upsilon\to\tau^+\tau^-$ decays (salmon line). The region above the white line labelled ``unphysical'' is forbidden for a given choice of $\tan\beta$. The differences with respect to Ref.~\cite{Babu:2017olk} show up in the high $M_{X}$ region.\label{fig:UV-correction}}
\end{figure*}

After symmetry breaking, a $\theta_W$ rotation in the  $(W^3,B)$ plane leads to a massless photon and  $Z_0$.
The latter mixes with the $X_0$ boson~\footnote{Here we call the gauge boson $X$ to be consistent with the notation of Ref.~\cite{Babu:2017olk}.} and the rotation that diagonalizes the gauge boson mass matrix is given by
\begin{equation}
    \sin(2\theta_{ZX}) = \frac{2}{3}g_{X}\frac{M_{Z}v\cos^2\beta}{M_Z^2-M_{X}^2}.
\end{equation}
Note that this is the exact expression for the mixing~\footnote{Note also that the masses $M_X$ and $M_Z$ are eigenvalues of a mass matrix, and thus are correlated, avoiding unphysical values for $\sin\theta_{XZ}$.}. 
The expression in Ref.~\cite{Babu:2017olk} is approximated and only valid for $M_{X}\ll M_{Z}$.
Some of the constraints derived in Ref.~\cite{Babu:2017olk} are thus valid only on that regime (others are unchanged). In Fig.~\ref{fig:UV-correction} we provide corrected low energy constraints on this UV complete scenario, and we slightly extend the $M_{X}$ range.

Finally, the shift in the electroweak $T$ parameter in this model is given by
\begin{equation}
    \Delta T\equiv \frac{1}{\alpha}\frac{\Delta M_Z^2}{M_Z^2} = \frac{1}{\alpha} \frac{\epsilon_{\tau\tau}}{3}\frac{M_X^2}{M_Z^2}\cos^2\beta.
\end{equation}
Note that for $M_X<M_Z$,  $\epsilon_{\tau\tau}$ is positive and therefore $\Delta T$ is also positive.
Contributions from the scalar sector will also be positive and thus the constraint from $T$ is robust for $M_X<M_Z$. 
When $M_X > M_Z$, $\epsilon_{\tau\tau}$ changes sign due to the change of sign in the mixing angle.
Thus, the contributions from the gauge sector can be compensated by those from the scalar sector, and the $T$ parameter constraint becomes more model dependent.

\begin{acknowledgments}
 The work of KB and SJ was supported in part by US Department of Energy Grant Number DE-SC 0016013 and  by the Neutrino Theory Network Program Grant No. DE-AC02-07CHI11359. KB, DG and SJ thank the Fermilab theory group for warm hospitality during the completion of this work. KB is supported in part by a Fermilab Distinguished Scholar program.  Fermilab is operated by the Fermi Research Alliance, LLC under contract No. DE-AC02-07CH11359 with the United States Department of Energy. 
 \end{acknowledgments}

\bibliographystyle{utphys}
\bibliography{reference}

\end{document}